\documentclass[%
 aip,
 jcp,                  
 amsmath,amssymb,
 reprint,%
]{revtex4-1}

\usepackage{graphicx}
\usepackage{dcolumn}
\usepackage{bm}
\usepackage{verbatim}
\usepackage{amsmath}

\begin{document}

\title[Van der Waals Interaction in Uniaxial Anisotropic Media]{Van der Waals Interaction in Uniaxial Anisotropic Media}

\author{Pavel Kornilovitch}
 \email{pavel.kornilovich@hp.com}
 \affiliation{HP Inc., Corvallis, Oregon 97330 USA} 


\date{\today}  

\begin{abstract}

Van der Waals interactions between flat surfaces in uniaxial anisotropic media are investigated in the nonretarded limit. The main focus is the effect of nonzero tilt between the optical axis and the surface normal on the strength of van der Waals attraction. General expressions for the van der Waals free energy are derived using the surface mode method and the transfer-matrix formalism. To facilitate numerical calculations a temperature-dependent three-band parameterization of the dielectric tensor of the liquid crystal 5CB is developed. A solid slab immersed in a liquid crystal experiences a van der Waals torque that aligns the surface normal relative to the optical axis of the medium. The preferred orientation is different for different materials. Two solid slabs in close proximity experience a van der Waals attraction that is strongest for homeotropic alignment of the intervening liquid crystal for all the materials studied. The results have implications for the stability of plate-like colloids in liquid crystal hosts.       

\end{abstract}

\pacs{61.30.-v}                      

\maketitle

\section{\label{sec:one}
Introduction
}

One component of the interaction between colloidal particles suspended in a fluid is van der Waals (vdW), or dispersion, forces~\cite{Verwey1999}. In the nonretarded limit (distances below approximately 100 nm) the dispersion forces are mediated by longitudinal matter-like modes localized on the interfaces, while in the retarded limit (large distances) they are mediated by transverse light-like modes standing between the interfaces~\cite{Sernelius2001}. Quantum-mechanical treatment of the nonretarded regime was pioneered by London~\cite{London1930} and of the retarded regime by Casimir and Polder~\cite{Casimir1948}. Lifshitz~\cite{Lifshitz1955} introduced a macroscopic treatment of the problem where the electromagnetic properties of the bodies and the medium were characterized through dielectric permittivity functions. Dzyaloshinskii and Pitaevskii generalized the Lifshitz theory to non-uniform bodies using the formalism of temperature Green's funstions~\cite{Dzyaloshinski1959,Dzyaloshinski1960,Dzyaloshinski1961}. Later, van Kampen, Nijboer, and Schram developed a method of calculating the vdW interaction based on the knowledge of electromagnetic surface states, which simplified the treatment of complex geometries~\cite{vanKampen1968}. Those pioneering works started theoretical and experimental investigation of vdW forces, a research field that has remained active ever since~\cite{Langbein1974,Barash1975,Mahanty1976,Parsegian2006,French2010}.

The present article concerns with vdW forces between nanosize particles in anisotropic media. It is motivated by experiments on colloidal suspensions in liquid crystals (LC). A variety of one-dimensional and two-dimensional self-assembled structures have been observed in suspensions of spherical particles in nematic LC~\cite{Poulin1997,Nazarenko2001,Misevic2006,Lapointe2009,Ognysta2011}. In a different experiment, stacking of plate-like clay particles in 5CB has been revealed by small-angle X-ray scattering in \cite{Pizzey2004,vanDuijneveldt2005,Connolly2006,Connolly2007}. Apart from fundamental interest, suspensions of colloids in nonaqueous solutions are of increasing technological importance, for instance as the work medium in electrophoretic reflective displays.  

In an isotropic host fluid, colloidal stability is largely determined by the balance between electrostatic and vdW forces. In an ordered LC, a third force emerges that originates from distortions of the director field. This elastic force is long-range, and in some situations completely dominates the force balance. For example, under the conditions of strong anchoring the elastic interaction energy between two micron-size {\em spherical} particles can reach 1000s of $kT$ \cite{Misevic2006,Ognysta2011}. The situation can be quite different for plate-shape particles for two reasons. First, a flat surface distorts the director field much less than a spherical surface even under strong anchoring. (Weak anchoring would further reduce the elastic energy.) Second, separation between two flat surfaces can be {\em uniformly} small, which boosts the vdW interaction. A typical value of the Hamaker constant is 10 zJ = $10^{-20}$ J $\approx$ 2 $kT$, see section~\ref{sec:four}. This yields for the vdW energy $F = (A/12 \pi)(D/L)^2 \approx (kT/20)(D/L)^2$, where $D$ is the plate diameter and $L$ the gap between the plates. As $L$ can be as small as a few nanometers \cite{Pizzey2004} aspect ratios $D/L \approx 10-100$ are possible, leading to vdW energies on the order of $(5-500) kT$. This simple estimate demonstrates that vdW can be a contributing factor in the overall energy balance.    

It is not the purpose of the present paper to develop a general theory of colloidal interaction in an anisotropic medium with full inclusion of vdW forces. That would be a complex undertaking as the director field affects the spectrum of the surface modes while the electrostatic field distorts the director. At the same time it is important to point out that when vdW forces {\em are} strong, they cannot be treated as isotropic. Instead, they acquire orientation dependence caused by the optical anisotropy of the host medium.     

The dielectric anisotropy was introduced into the vdW problem, apparently for the first time, by Kihara and Honda~\cite{Kihara1965}. Those authors considered electrically anisotropic solid particles separated by an isotropic liquid. Kats~\cite{Kats1971} generalized the temperature Green's function formalism to anisotropic media. He considered two anisotropic particles separated by an isotropic liquid and calculated a vdW {\em torque} that rotates the two particles relative to each other. Kats also considered a cholesteric liquid crystal mediating vdW interaction between two isotropic bodies. Parsegian and Weiss~\cite{Parsegian1972} studied two anisotropic bodies interacting across an {\em anisotropic medium} but with all three regions sharing one common optical axis. Barash~\cite{Barash1978} generalized their results to include the retardation effects. Smith and Ninham~\cite{Smith1973} investigated the vdW forces applied by two anisotropic bodies on a film of twisted nematic squeezed between them. The torque exerted by the bodies was balanced by the elastic energy of the nematic, which determined the structure of the twist. Munday et al~\cite{Munday2005} calculated the torque and force between two anisotropic disks across an isotropic fluid and found parameters when the force would be repulsive. Under those conditions the disk could levitate over the substrate and the torque on the disk could become measurable. \v{S}arlah and \v{Z}umer~\cite{Sarlah2001} considered two semi-infinite optically uniaxial bodies separated by a uniaxial medium with all three optical axes parallel. Those authors derived an analytical expression for the Hamaker constant but did not consider vdW torques. Veble and Podgornik~\cite{Veble2009} developed the general theoretical framework for layered anisotropic materials in terms of $4 \times 4$ transfer matrices.  

The main goal of the present paper is to investigate the effects of arbitrary orientation of the optical axis relative to particle surfaces. Due to the complexity of the general problem, treatment will be confined to the parallel-plate geometry in the nonretarded limit. In this case, the vdW energy and forces can be derived from the spectrum of longitudinal surface modes~\cite{vanKampen1968,Mahanty1976,Parsegian2006} but the general formalism of \cite{Veble2009} is not needed. The electrostatic model is formulated in section~\ref{sec:two} and the general solution is constructed in section~\ref{sec:three}. For numerical evaluation of forces dielectric functions at imaginary frequencies are needed. Such functions for several materials are listed in section~\ref{sec:five}. In addition, a temperature-dependent dielectric model of 5CB is defined for both polarizations. The single-slab problem is solved in section~\ref{sec:seven} and the two-slab problem in section~\ref{sec:four}.

\begin{figure}[t]
\begin{center}
\includegraphics[width=0.48\textwidth]{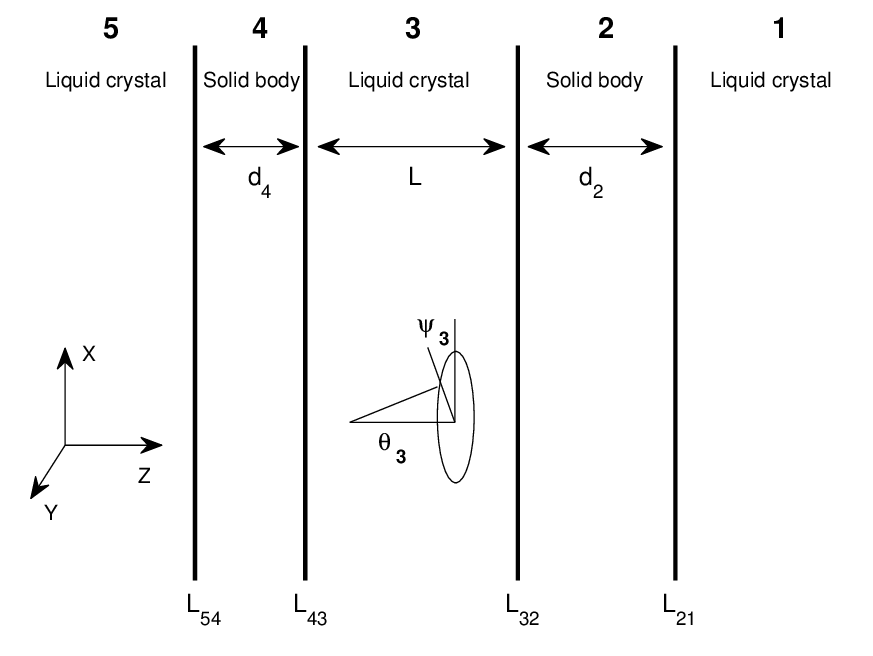}
\end{center}
\caption{The model geometry. Two finite thickness slabs {\bf 2} and {\bf 4} made of dielectrically isotropic materials are immersed in uniaxial liquid crystals {\bf 1}, {\bf 3}, and {\bf 5}. The slabs are infinite in the $x$ and $y$ directions and parallel to each other. The optical axes of the liquid crystals are tilted away from the surface normal by respective angles $\theta_{1,3,5}$ and rotated around the $z$-axis by respective angles $\psi_{1,3,5}$.}
\label{fig:one}
\end{figure}

\section{\label{sec:two}
Model and method
}

The overall geometry is shown in figure~\ref{fig:one}. The system of interest consists of two parallel slabs {\bf 2} and {\bf 4} a distance $L \equiv L_{32}-L_{43}$ apart. All slabs' surfaces are parallel to the $xy$-plane of the coordinate system, and the $z$-axis is perpendicular to the interfaces. The slabs are infinite in the $x$ and $y$ directions and have finite thicknesses in the $z$ direction: $d_2$ and $d_4$, respectively. The materials of {\bf 2} and {\bf 4} are assumed to be optically and dielectrically isotropic. They are characterized by the scalar dielectric functions $\varepsilon_{\bf 2}(\omega)$ and $\varepsilon_{\bf 4}(\omega)$. [The formalism developed below could be easily generalized to anisotropic solid materials. However, it would have obscured the main focus of the paper, which are the effects derived from the anisotropy of the medium. Such a generalization is left for future research.]         

The space between the slabs as well as outside the slabs is filled with dielectrically anisotropic media. The respective regions are labeled ${\bf r} = {\bf 1}$, {\bf 3}, and {\bf 5}. The media are assumed to be optically uniaxial such as most liquid crystals. (The terms ``uniaxial media'' and ``liquid crystals'' will be used interchangeably throughout the paper.) The dielectric and optical properties are characterized by the parallel and perpendicular dielectric functions $\varepsilon^{\parallel}_{\bf r}(\omega,T)$ and $\varepsilon^{\perp}_{\bf r}(\omega,T)$. In uniaxial liquid crystals, both dielectric functions are strong functions of temperature $T$. The theoretical formalism allows the three materials to be different. However in all the examples considered below the material will be the same in all three regions.  

In each region {\bf r}, the direction of the optical axis is defined by polar angle $\theta_{r}$ and azimuth angle $\psi_{r}$, as shown in figure~\ref{fig:one}. The dielectric tensor is defined as follows. (i) In the starting orientation the optical axis coincides with the $z$-axis. This implies $\varepsilon^{zz} = \varepsilon^{\parallel}$, $\varepsilon^{xx} = \varepsilon^{yy} = \varepsilon^{\perp}$, and all off-diagonal terms are zero. (ii) The medium is rotated around axis $y$ by an angle $\theta$. (iii) The medium is rotated around axis $z$ by an angle $\psi$. The resulting dielectric tensor in region {\bf r} is 
\begin{equation}
\hat{\varepsilon}^{jk}_{\bf r}(\theta_r,\psi_r,\omega,T) = 
\left[ 
\begin{array}{ccc}
\varepsilon^{xx}_{\bf r} & \varepsilon^{xy}_{\bf r} & \varepsilon^{xz}_{\bf r}  \\
\varepsilon^{yx}_{\bf r} & \varepsilon^{yy}_{\bf r} & \varepsilon^{yz}_{\bf r}  \\
\varepsilon^{zx}_{\bf r} & \varepsilon^{zy}_{\bf r} & \varepsilon^{zz}_{\bf r}   
\end{array} \right] \! ,
\label{eq:one}
\end{equation}
where
\begin{eqnarray}
\varepsilon^{xx}_{\bf r} & = & 
\varepsilon^{\perp}_{\bf r}     ( \cos^2{\theta_r} \cos^2{\psi_r} + \sin^2{\psi_r} ) + 
\varepsilon^{\parallel}_{\bf r} ( \sin^2{\theta_r} \cos^2{\psi_r} ) 
\nonumber    \\ 
\varepsilon^{xy}_{\bf r} & = & \varepsilon^{yx}_{\bf r} = 
( \varepsilon^{\parallel}_{\bf r} - \varepsilon^{\perp}_{\bf r} ) 
\sin^2{\theta_r} \sin{\psi_r} \cos{\psi_r}
\nonumber    \\
\varepsilon^{xz}_{\bf r} & = & \varepsilon^{zx}_{\bf r} = 
( \varepsilon^{\perp}_{\bf r} - \varepsilon^{\parallel}_{\bf r} ) 
\sin{\theta_r} \cos{\theta_r} \cos{\psi_r}
\nonumber    \\
\varepsilon^{yy}_{\bf r} & = & 
\varepsilon^{\perp}_{\bf r}     ( \cos^2{\theta_r} \sin^2{\psi_r} + \cos^2{\psi_r} ) + 
\varepsilon^{\parallel}_{\bf r} ( \sin^2{\theta_r} \sin^2{\psi_r} ) 
\nonumber    \\
\varepsilon^{yz}_{\bf r} & = & \varepsilon^{zy}_{\bf r} = 
( \varepsilon^{\perp}_{\bf r} - \varepsilon^{\parallel}_{\bf r} ) 
\sin{\theta_r} \cos{\theta_r} \sin{\psi_r}
\nonumber    \\
\varepsilon^{zz}_{\bf r} & = & 
\varepsilon^{\perp}_{\bf r} \sin^2{\theta_r} + 
\varepsilon^{\parallel}_{\bf r} \cos^2{\theta_r}  \: . 
\label{eq:two}
\end{eqnarray}

All five regions are assumed to be spatially uniform, i.e. the dielectric functions are independent of $x$, $y$, and $z$. This implies no distortion of the director field and hence a zero bulk elastic energy. Physically, this corresponds to either very weak or very strong anchoring. In the first case, the surface energy is zero, and the orientation is an arbitrary parameter. In the second case, the orientation is fixed and the vdW energy is a small correction to the dominant surface energy. In any case, the goal of the present calculation is to characterize the vdW energy for a given orientation which is considered a model parameter.          

In the surface mode method~\cite{vanKampen1968,Barash1975,Mahanty1976,Parsegian2006}, the vdW energy is derived from the spectrum of electromagnetic surface modes. If $W(q,\omega) = 0$ is a spectral equation, then the vdW free energy is a sum over all quantum numbers $q$ and bosonic imaginary frequencies $\xi_n = (2 \pi k T/\hbar) n$ 
\begin{equation}
F = kT \sum_q \sideset{}{'}\sum^{\infty}_{n = 0} \ln \left[ W(q, {\rm i} \xi_n) \right] \: . 
\label{eq:three}
\end{equation}
The prime at the sum sign indicates the $n = 0$ term must be taken with weight $\frac{1}{2}$. Differentiating with respect to the distance between the bodies yields the interaction force.  

The method is particularly simple in the non-retarded limit when surface modes are obtained from solutions of the anisotropic Laplace equation rather than of the full set of Maxwell's equations:
\begin{equation}
\frac{\partial}{\partial x_j} \left( \varepsilon^{jk} \frac{\partial \phi}{\partial x_k} \right) = 0 \: . 
\label{eq:four}
\end{equation}
The modes are fixed by appropriate boundary conditions at phase boundaries. The solution can be constructed by using transfer matrices. This will be done in section~\ref{sec:three}.

\section{\label{sec:five}
Material properties
}

Numerical evaluation of the vdW energy and forces requires dielectric functions of the interacting materials. Accurate knowledge of the entire tensor $\hat{\varepsilon}(\omega,T)$ is of paramount importance. Before solving the electrostatic surface mode problem, material properties are discussed in this section.    

For a growing number of substances, the dielectric functions on the imaginary axis are obtained through a Kramers-Kronig transformation of the absorption or reflection data followed by a fit to a multiple Lorentz oscillator model~\cite{Mahanty1976}. In this paper, the parameters given by Parsegian~\cite{Parsegian2006} and van Zwol and Palasantzas~\cite{vanZwol2010} are used to construct dielectric functions of the following isotropic materials: silica~\cite{vanZwol2010} (set 1); polytetrafluoroethylene~\cite{vanZwol2010} (Teflon or PTFE); polystyrene~\cite{vanZwol2010} (set 2), mica~\cite{Parsegian2006} (Table L2.7, set $b$), gold~\cite{Parsegian2006} (Table L2.4, set 3), silver~\cite{Parsegian2006} (Table L2.5, set 1), and copper~\cite{Parsegian2006} (Table L2.6).

\begin{table}[b]
\renewcommand{\tabcolsep}{0.2cm}
\renewcommand{\arraystretch}{1.5}
\begin{center}
\begin{tabular}{|c|c|c|c|c|}
\hline\hline
  $T$, (K)       &  $C_{1e}$  &  $C_{2e}$  &  $C_{1o}$  &  $C_{2o}$   \\ \hline\hline
   298.2         &  0.10253   &  0.11110   &  0.05960   &  0.03737    \\ \hline
   300.3         &  0.09724   &  0.11027   &  0.06011   &  0.03808    \\ \hline
   303.0         &  0.09205   &  0.10771   &  0.06104   &  0.03962    \\ \hline
   305.7         &  0.08470   &  0.10535   &  0.06234   &  0.04193    \\ \hline
   307.9         &  0.07587   &  0.09071   &  0.06427   &  0.04631    \\ \hline
\hline \end{tabular}
\end{center}
\caption{Temperature-dependent coefficients $C$ of the three-band di\-e\-lec\-tric model, (\ref{eq:sevenone})-(\ref{eq:seventwo}), of 4-cyano-4-$n$-pen\-tyl\-bi\-phe\-nyl (5CB).~\cite{Wu1993} Other model parameters are: $\omega_0 = 9.19$ eV, $\omega_1 = 5.91$ eV, $\omega_2 = 4.40$ eV, $T_c = 308.3$ K, and $\beta = 0.142$. }
\label{tab:one}
\end{table}
\begin{figure*}[t]
\begin{center}
\includegraphics[width=0.98\textwidth]{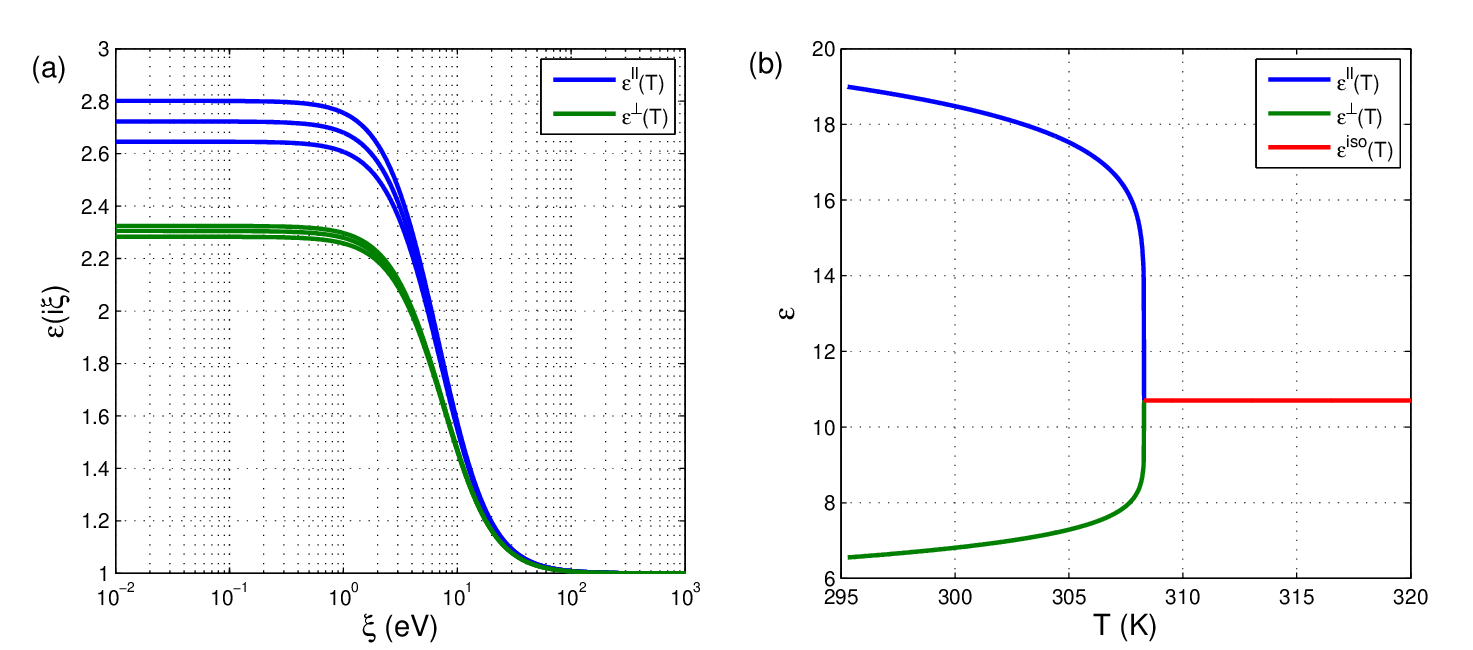}
\end{center}
\caption{(a) Dynamic dielectric functions of liquid crystal 5CB, (\ref{eq:sevenone})-(\ref{eq:seventwo}) for the three temperatures 298.2 K, 305.7 K, and 307.9 K. $\varepsilon^{\parallel}$ decreases with temperature while $\varepsilon^{\perp}$ increases with temperature. (b) Static dielectric functions of 5CB, (\ref{eq:fiveseven})-(\ref{eq:fiveeight}).}
\label{fig:oneone}
\end{figure*}

There is much less information on $\hat{\varepsilon}({\rm i}\xi,T)$ of liquid crystals. The usual difficulty of knowing the optical spectra in a wide energy interval is multiplied here by the need to know them separately for two polarizations and at different temperatures. Only a few liquid crystals have been studied experimentally well enough to enable a full model. One of the most studied compounds is 4-cyano-4-$n$-pentylbiphenyl (5CB), which will be used here as an  exemplary positive uniaxial material.  

The dielectric functions of 5CB used in this paper are based on the three-band dispersion model developed by Wu and co-workers~\cite{Wu1990,Wu1991,Wu1993,Li2004}. The model accurately describes the experimentally measured refractive indices in the (0.4-0.8) $\mu$m spectral interval for the entire temperature interval of the nematic phase 295.3 K $\leq T \leq$ 308.3 K. According to the model, the index dispersion in the visible is governed by three electronic transitions: one $\sigma \rightarrow \sigma^{\ast}$ transition with $\lambda_0 \approx 0.135$ $\mu$m, and two $\pi \rightarrow \pi^{\ast}$ transitions with $\lambda_1 = 0.210$ $\mu$m and $\lambda_2 = 0.282$ $\mu$m. The oscillator strength of the $\sigma \rightarrow \sigma^{\ast}$ transition is very weakly temperature dependent. It can be extracted from the dispersion of the isotropic part of the refractive index. The oscillator strengths of the $\pi \rightarrow \pi^{\ast}$ transitions are temperature-dependent. In particular, they define the temperature variation of birefringence. Converting to imaginary frequency and using $\varepsilon^{\parallel, \perp}({\rm i}\xi,T) = n^2_{e,o}({\rm i}\xi,T)$, the model reads 
\begin{eqnarray}
\varepsilon^{\parallel}_{\rm 5CB}({\rm i}\xi,T) & = & 
\left[ 1 + \frac{0.460}{1+\frac{\xi^2}{\omega^2_0}} + 
           \frac{C_{1e}(T)}{1+\frac{\xi^2}{\omega^2_1}} + 
           \frac{C_{2e}(T)}{1+\frac{\xi^2}{\omega^2_2}} \right]^2 \!\! ,
\label{eq:sevenone} \\ 
\varepsilon^{\perp}_{\rm 5CB}({\rm i}\xi,T) & = & 
\left[ 1 + \frac{0.414}{1+\frac{\xi^2}{\omega^2_0}} + 
           \frac{C_{1o}(T)}{1+\frac{\xi^2}{\omega^2_1}} + 
           \frac{C_{2o}(T)}{1+\frac{\xi^2}{\omega^2_2}} \right]^2 \!\! .
\label{eq:seventwo}
\end{eqnarray}
Here $\omega_0 = 9.19$ eV, $\omega_1 = 5.91$ eV, and $\omega_2 = 4.40$ eV. The refractive indices of 5CB were measured by polarized UV spectroscopy~\cite{Wu1990} and tabulated by Wu et al~\cite{Wu1993}. The temperature-dependent coefficients $C$ extracted from those data are given in table~\ref{tab:one}.  

Rotational relaxation of 5CB molecules and other low frequency processes are neglected here based on the familiar argument~\cite{Sarlah2001} that their characteristic energies of less than $0.01$ eV are much smaller than the first bosonic frequency at room temperature $\xi_1 \approx 0.16$ eV. Equations (\ref{eq:sevenone})-(\ref{eq:seventwo}) represent the entire {\em dynamical} part of the dielectric functions. The functions are plotted in figure~\ref{fig:oneone}(a). 

Static response requires a separate treatment. The static dielectric constant can be split into a temperature-independent isotropic part ($\approx 10.7$ for 5CB) and a temperature-dependent birefringent part. It is reasonable to assume that the temperature dependence comes from the order parameter. According to Li and Wu~\cite{Li2004} the order parameter of 5CB follows a universal relation $\propto (1 - T/T_c)^{\beta}$, where $T_c = 308.3$ is the nematic-isotropic transition temperature and $\beta = 0.142$ is a universal exponent. Adjusting the dielectric constants to the measured experimental values at low temperatures~\cite{Chandrasekhar1977}, one arrives at the following parameterization       
\begin{eqnarray}
\varepsilon^{\parallel}_{\rm 5CB}(0,T) & = & 10.7 + 13.0 \left( 1 - \frac{T}{T_c} \right)^{\beta} ,
\label{eq:fiveseven} \\ 
\varepsilon^{\perp}_{\rm 5CB}(0,T)     & = & 10.7 -  6.5 \left( 1 - \frac{T}{T_c} \right)^{\beta} , 
\label{eq:fiveeight}
\end{eqnarray}
with $T_c = 308.3$ K and $\beta = 0.142$. These functions are plotted in figure~\ref{fig:oneone}(b).

Several comments are now in order. (i) The dynamical model (\ref{eq:sevenone})-(\ref{eq:seventwo}) is defined only at 5 discrete temperatures listed in table~\ref{tab:one}. The model can be extended to any temperature by fitting the coefficients to the same universal factor $a + b(1 - T/T_c)^{\beta}$~\cite{Li2004}. This is not done in the present work. (ii) It is possible to convert (\ref{eq:sevenone})-(\ref{eq:seventwo}) to a more familiar form by expanding the square and refitting the function to a linear combination of oscillators. Such an additional fitting procedure might introduce undesirable errors and therefore is not employed here. (iii) Wu et al~\cite{Wu1993} provided a data set for another liquid crystal compound 5PCH, thus enabling a similar three-band dielectric model.

\section{\label{sec:three}
Transfer matrices
}

The anisotropic Laplace equation (\ref{eq:four}) is now solved. In analogy with the optics of multilayered media, it is convenient to construct solutions out of individual transfer matrices~\cite{Parsegian2006,Veble2009}. Each matrix links wave amplitudes on either side of one interface, and the full amplitude is found by matrix multiplications. The two fundamental transfer matrices are derived below.    

\begin{widetext}

The general solution in each region is a linear combination of the transmitted and reflected waves. Consider interface ({\bf 21}). The solution in medium region {\bf 1} is sought in the form
\begin{equation}
\phi_{\bf 1} = \left[ A_{\bf 1} \: 
\exp{\left( \frac{\lambda_{\bf 1}^{+}}{\varepsilon^{zz}_{\bf 1}} z \right)} 
+ B_{\bf 1} \: 
\exp{\left( \frac{\lambda_{\bf 1}^{-}}{\varepsilon^{zz}_{\bf 1}} z \right)}  \right]
\exp{[{\rm i}(q_x x + q_y y)]}  \: .
\label{eq:five}
\end{equation}
Here $A_{\bf 1}$ and $B_{\bf 1}$ are unknown amplitudes, $\varepsilon^{zz}_{\bf 1}$ in the exponent is introduced for convenience, and the last exponential factor reflects the uniformity of the Laplace equation in the $(xy)$ plane. The exponents $\lambda^{\pm}_{\bf 1}$ follow from (\ref{eq:four}) and (\ref{eq:two}): 
\begin{equation}
\lambda^{\pm}_{\bf 1} = -{\rm i} 
( q_x \varepsilon^{xz}_{\bf 1} + q_y \varepsilon^{yz}_{\bf 1} ) \pm p_{\bf 1} \: , 
\label{eq:six}
\end{equation}
\begin{eqnarray}
p^2_{\bf 1} & \equiv & q^2_x \left( \varepsilon^{\perp}_{\bf 1} \varepsilon^{\parallel}_{\bf 1} \right) 
                             \left( \cos^2{\theta_1} \sin^2{\psi_1} + \cos^2{\psi_1} \right) 
                 +     q^2_x \left( \varepsilon^{\perp}_{\bf 1} \right)^2 
                             \left( \sin^2{\theta_1} \sin^2{\psi_1} \right)        
\nonumber \\
            &   +    & 2 q_x q_y \: \varepsilon^{\perp}_{\bf 1} 
                          \left( \varepsilon^{\parallel}_{\bf 1} - \varepsilon^{\perp}_{\bf 1} \right) 
                          \left( \sin^2{\theta_1} \cos{\psi_1} \sin{\psi_1} \right)                    
                +      q^2_y \left( \varepsilon^{\perp}_{\bf 1} \varepsilon^{\parallel}_{\bf 1} \right) 
                             \left( \cos^2{\theta_1} \cos^2{\psi_1} + \sin^2{\psi_1} \right)   
\nonumber \\                             
            &   +    & q^2_y \left( \varepsilon^{\perp}_{\bf 1} \right)^2 
                             \left( \sin^2{\theta_1} \cos^2{\psi_1} \right)        \: . 
\label{eq:sixone}
\end{eqnarray}
The quantity $p$ is defined as the positive square root of $p^2$. $p$ is a function of the interface momentum components $q_x$ and $q_y$, optical axis angles $\theta$ and $\psi$, imaginary frequency $\xi$, and temperature $T$. In slab region {\bf 2} the material is isotropic and the Laplace equation simply yields 
\begin{equation}
\phi_{\bf 2} = \left( A_{\bf 2} \: e^{q z} + B_{\bf 2} \: e^{- q z} \right) 
\exp{[{\rm i}(q_x x + q_y y)]} \: ,  
\label{eq:seven}
\end{equation}
\begin{equation}
q \equiv + \sqrt{q^2_x + q^2_y} \: . 
\label{eq:eight}
\end{equation}
The matching conditions at $z = L_{21}$ include the equality of the transverse components of the electric field $E^x = -\partial \phi/\partial x$ and $E^y = -\partial \phi/\partial y$ (which lead to identical relationships), and the equality of the normal components of $D^z = - \varepsilon^{zk} (\partial \phi/\partial x_k)$. The resulting two equations can be rearranged to express the wave amplitudes of region {\bf 2} via the wave amplitudes of region {\bf 1}:     
\begin{equation}
\left( \begin{array}{c}  A_{\bf 2} \\  B_{\bf 2}  \end{array} \right) = 
\hat{M}_{\bf 21} \left( \begin{array}{c} A_{\bf 1} \\  B_{\bf 1} \end{array} \right) . 
\label{eq:eightone}
\end{equation}
\begin{equation}
\hat{M}_{\bf 21} = 
\left\{ \begin{array}{cc} 
{\displaystyle \frac{q \varepsilon_{\bf 2} + p_{\bf 1}}{2 \, q \varepsilon_{\bf 2}} } \:  
\exp{\left[ \left( - q + \frac{\lambda_{\bf 1}^{+}}{\varepsilon^{zz}_{\bf 1}} \right) L_{21} \right]}  
 & {\displaystyle \frac{q \varepsilon_{\bf 2} - p_{\bf 1}}{2 \, q \varepsilon_{\bf 2}} } \:
\exp{\left[ \left( - q + \frac{\lambda_{\bf 1}^{-}}{\varepsilon^{zz}_{\bf 1}} \right) L_{21} \right]} \\  
{\displaystyle \frac{q \varepsilon_{\bf 2} - p_{\bf 1}}{2 \, q \varepsilon_{\bf 2}} } 
\exp{\left[ \left(   q + \frac{\lambda_{\bf 1}^{+}}{\varepsilon^{zz}_{\bf 1}} \right) L_{21} \right]} 
 & {\displaystyle \frac{q \varepsilon_{\bf 2} + p_{\bf 1}}{2 \, q \varepsilon_{\bf 2}} } \:  
\exp{\left[ \left(   q + \frac{\lambda_{\bf 1}^{-}}{\varepsilon^{zz}_{\bf 1}} \right) L_{21} \right]} 
\end{array}\right\} .
\label{eq:nine}
\end{equation}
The last equality defines the transfer matrix $\hat{M}_{\bf 21}$ between the medium region {\bf 1} and slab region {\bf 2}. A similar transfer matrix defines scattering at the ({\bf 34}) interface after index substitution ${\bf 1} \rightarrow {\bf 3}$ and ${\bf 2} \rightarrow {\bf 4}$.  At interface ({\bf 23}), the waves are incident from slab region {\bf 2}. From the matching conditions one expresses the medium amplitudes $A_{\bf 3}$ and $B_{\bf 3}$ via the slab amplitudes $A_{\bf 2}$ and $B_{\bf 2}$. After some algebra one obtains
\begin{equation}
\left( \begin{array}{c} A_{\bf 3} \\  B_{\bf 3}  \end{array} \right) = 
\hat{M}_{\bf 32} \left( \begin{array}{c} A_{\bf 2} \\  B_{\bf 2} \end{array} \right) . 
\label{nineone}
\end{equation}
\begin{equation}
\hat{M}_{\bf 32} = 
\left\{ \begin{array}{cc} 
{\displaystyle \frac{p_{\bf 3} + q \varepsilon_{\bf 2}}{2p_{\bf 3}} } \:  
\exp{\left[ \left(   q - \frac{\lambda_{\bf 3}^{+}}{\varepsilon^{zz}_{\bf 3}} \right) L_{32} \right] } 
 & {\displaystyle \frac{p_{\bf 3} - q \varepsilon_{\bf 2}}{2p_{\bf 3}} } \:
\exp{\left[ \left( - q - \frac{\lambda_{\bf 3}^{+}}{\varepsilon^{zz}_{\bf 3}} \right) L_{32} \right] } \\  
{\displaystyle \frac{p_{\bf 3} - q \varepsilon_{\bf 2}}{2p_{\bf 3}} } 
\exp{\left[ \left(   q - \frac{\lambda_{\bf 3}^{-}}{\varepsilon^{zz}_{\bf 3}} \right) L_{32} \right] } 
 & {\displaystyle \frac{p_{\bf 3} + q \varepsilon_{\bf 2}}{2p_{\bf 3}} } \:  
\exp{\left[ \left( - q - \frac{\lambda_{\bf 3}^{-}}{\varepsilon^{zz}_{\bf 3}} \right) L_{32} \right] } 
\end{array}\right\} .
\label{eq:ten}
\end{equation}
A similar matrix describes scattering at the ({\bf 54}) interface. 
\end{widetext}
\begin{figure*}[t]
\begin{center}
\includegraphics[width=0.98\textwidth]{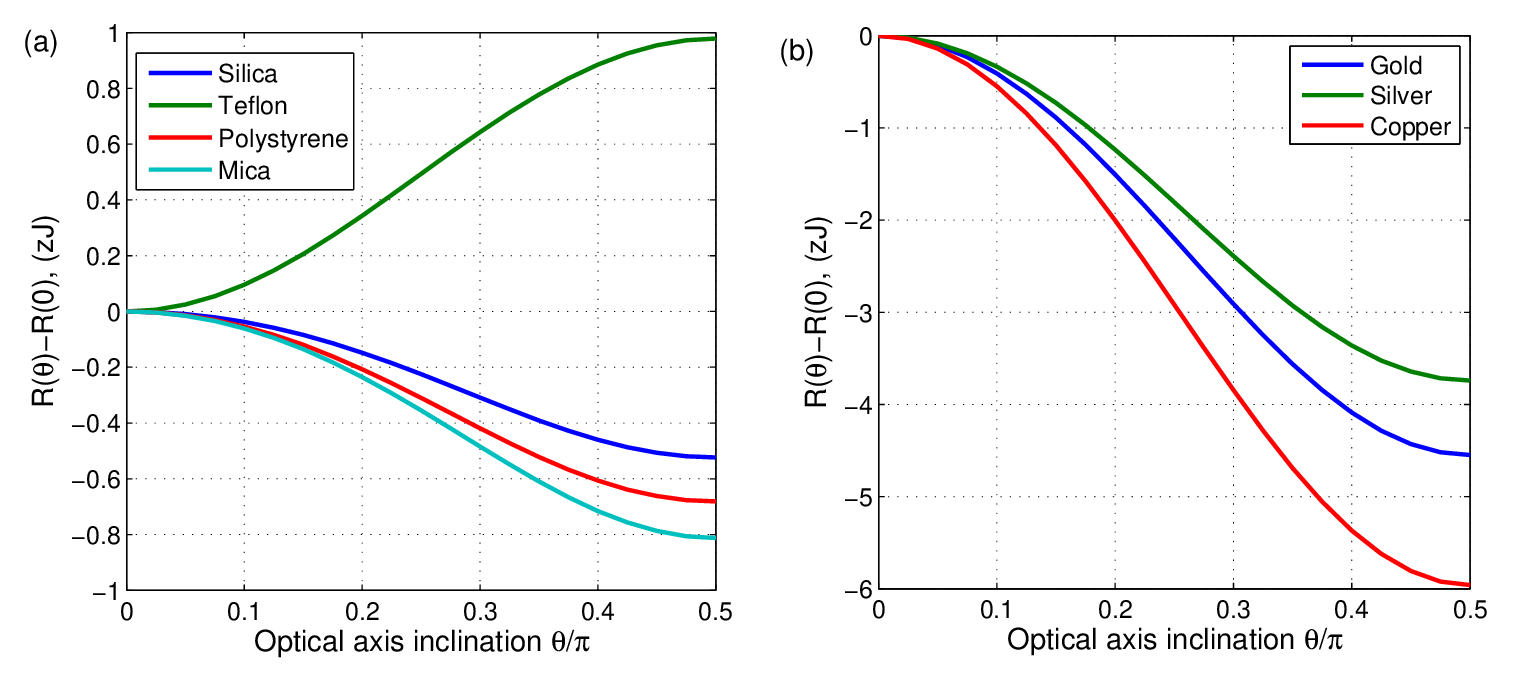}
\end{center}
\caption{(a) Tilt Hamaker constant (\ref{eq:sixtythreeone}) of a parallel-plate solid slab immersed in 5CB, as a function of the optical axis tilt angle $\theta$. The tilt is the same on both sides, $\theta_1 = \theta_3$. The azimuth angles are $\psi_1 = \psi_3 = 0$. The $R$s are referenced from their respective values $R(0) = -14.40, \: -12.90, \: -9.21, \: -4.25$ zJ for silica, Teflon, polystyrene, and mica, respectively. Teflon favors homeotropic alignment, $\theta = 0$, while other materials favor the planar alignment, $\theta = \pi/2$. Absolute temperature is $T = 298.2$ K. (b) Same for several metals in 5CB. The references are $R(0) = -103.63$, $-116.13$, $-244.62$ zJ for gold, silver, and copper, respectively. All metals favor the planar alignment, $\theta = \pi/2$.}
\label{fig:oneseven}
\end{figure*}

\section{\label{sec:seven}
One slab
}

In an isotropic liquid, a parallel-plate slab does not experience any macroscopic forces or torques. In an anisotropic liquid, the dependence of the dispersion forces on the inclination angle will result in a torque acting on the slab. For planar and other non-homeotropic surface alignments, the orientation of the optical axes on the two sides of the slab may in principle be different and this effect also warrants analysis. If, in addition, the solid material itself is anisotropic, there will be another torque that will rotate the plate around its normal. The latter effect is not considered in this paper. (Note that all the cases mentioned are different from the mutual torque between {\em two} anisotropic bodies studied by Kats~\cite{Kats1971} and Parsegian and Weiss~\cite{Parsegian1972}).

To determine the vdW energy of a single slab in an anisotropic host, only three regions of figure~\ref{fig:one} need to be taken into account, for instance {\bf 1}, {\bf 2}, and {\bf 3}. The scattering problem involves two transfer matrices
\begin{equation}
\left( \begin{array}{c} A_{\bf 3} \\  B_{\bf 3}  \end{array} \right) = 
\hat{M}_{\bf 32} \cdot \hat{M}_{\bf 21} 
\left( \begin{array}{c} A_{\bf 1} \\  B_{\bf 1} \end{array} \right) . 
\label{eq:sixty}
\end{equation}
Surface states are defined as exponentially decaying at infinity. Accordingly, the amplitudes $A_{\bf 1}$ and $B_{\bf 3}$ must be set to zero. The top equation (\ref{eq:sixty}) links the wave amplitudes $A_{\bf 3}$ and $B_{\bf 1}$ on either side of the system and hence defines the spatial structure of the surface mode. The bottom equation has the form $W \cdot B_{\bf 1} = 0$. For a non-vanishing $B_{\bf 1}$, this implies $W = 0$, which yields the surface mode spectrum. Expressed via matrix elements of the transfer matrices, the spectrum equation is
\begin{equation}
M_{\bf 32}^{21} M_{\bf 21}^{12} + M_{\bf 32}^{22} M_{\bf 21}^{22} = 0 \: . 
\label{eq:sixtyone}
\end{equation}
Substituting the matrix elements from (\ref{eq:nine}) and (\ref{eq:ten}) it becomes
\begin{equation}
W^{t} = 1 - 
\frac{ ( q \varepsilon_{\bf 2} - p_{\bf 3} )( q \varepsilon_{\bf 2} - p_{\bf 1} ) }
     { ( q \varepsilon_{\bf 2} + p_{\bf 3} )( q \varepsilon_{\bf 2} + p_{\bf 1} ) }  
\cdot \exp{ ( - q d_2 ) } = 0 \: . 
\label{eq:sixtytwo}
\end{equation}
The free energy is then obtained as follows: (i) The spectrum equation (\ref{eq:sixtytwo}) is substituted in (\ref{eq:three}); (ii) Polar coordinates $q_x = q \cos{\chi}$, $q_y = q \sin{\chi}$, are employed in the integral over the surface vector; (iii) A new function $u_{\bf r} \equiv p_{\bf r}/q$ is introduced. It is a function of the momentum angle $\chi$ but not of the momentum amplitude $q$. The explicit form of $u_{\bf r}$ follows from (\ref{eq:sixone}). (iv) The logarithm is expanded in an infinite series and integration over $q$ is performed. The resulting expression for the free energy {\em per unit area} is     
\begin{equation}
F^{t}_{1} = - \frac{kT}{8 \pi d^2_2} \sideset{}{'}\sum^{\infty}_{n = 0} 
\sum^{\infty}_{m=1} \frac{1}{m^3} \int^{2\pi}_{0} \! \frac{{\rm d}\chi}{2\pi} 
\left( \Delta_{\bf 23} \Delta_{\bf 21} \right)^{m} \: , 
\label{eq:sixtythree}
\end{equation}
where
\begin{equation}
\Delta_{\bf 23} \equiv \frac{ q \varepsilon_{\bf 2} - p_{\bf 3} }{ q \varepsilon_{\bf 2} + p_{\bf 3} }
                    =  \frac{   \varepsilon_{\bf 2} - u_{\bf 3} }{   \varepsilon_{\bf 2} + u_{\bf 3} }  \: , 
\label{eq:sixtyfour}
\end{equation}
and the same formula holds for $\Delta_{\bf 23}$ with {\bf 3} replaced by {\bf 1}. By analogy with vdW interaction between two semi-infinite bodies, the overall $1/d^{2}$ dependence can be isolated by introducing a Hamaker-like constant $R$ defined as $F^{t}_{1} = R/(12\pi d^2_{2})$. One arrives at
\begin{equation}
R(\theta,\psi,T) = - \frac{3kT}{2} \sideset{}{'}\sum^{\infty}_{n = 0}  
\sum^{\infty}_{m=1} \frac{1}{m^3} \int^{2\pi}_{0} \! \frac{{\rm d}\chi}{2\pi} 
\left( \Delta_{\bf 23} \Delta_{\bf 21} \right)^{m} . 
\label{eq:sixtythreeone}
\end{equation}
Note that $R$ is defined with a negative overall sign to retain the visual appeal of energy profiles: large negative $R$ implies lower energy and preferred orientation. In the following, $R$ will be referred to as ``tilt Hamaker constant'' to reflect its relation to the inclination of the optical axis.

\begin{figure}[t]
\begin{center}
\includegraphics[width=0.48\textwidth]{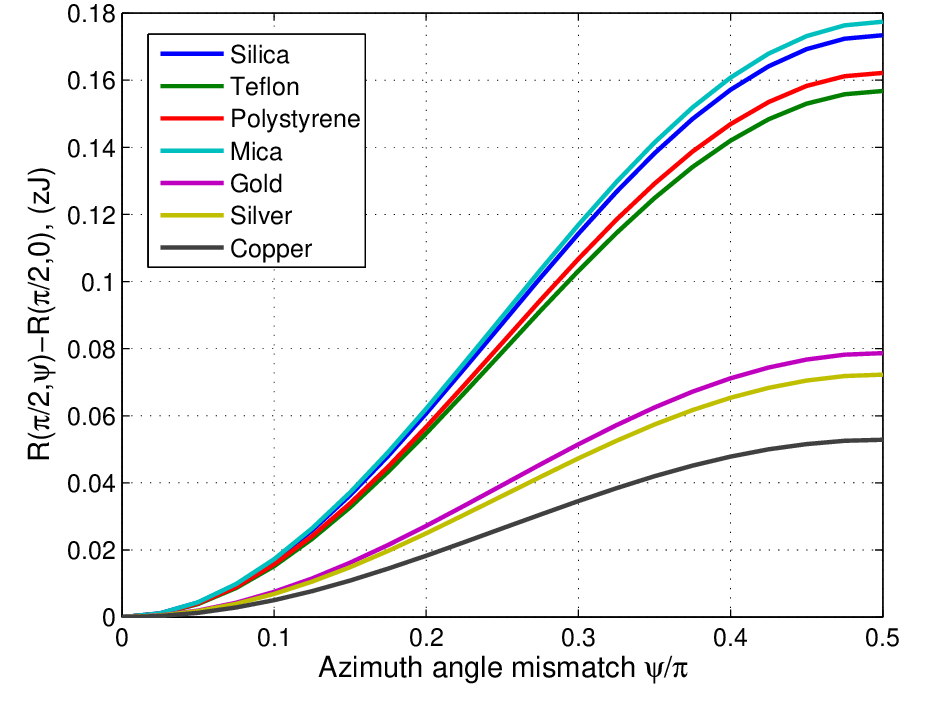}
\end{center}
\caption{Tilt Hamaker constant (\ref{eq:sixtythreeone}) for single slabs in 5CB. The alignment is planar on both sides of the slab, $\theta_1 = \theta_3 = \pi/2$ but the azimuth angle difference is systematically varied. The reference values are $R(\pi/2,0) = -14.93$, $-11.92$, $-9.90$, $-5.06$, $-108.18$, $-119.87$, $-250.58$ zJ for silica, Teflon, polystyrene, mica, gold, silver, and copper, respectively. All materials favor the parallel alignment of optical axes, $\psi_1 = \psi_3$ and do not favor the cross alignment $\psi_3 - \psi_1 = \pi/2$. Absolute temperature is $T = 298.2$ K.}
\label{fig:twozero}
\end{figure}

Figure~\ref{fig:oneseven}(a) shows tilt Hamaker constants for several dielectric materials immersed in 5CB. The tilt angle is the same on both sides of the slab, $\theta_1 = \theta_3$ and $\psi_1 = \psi_3 = 0$. For better presentation, the constants are referenced from their respective values at homeotropic alignment $\theta = 0$. The reference values are listed in the caption. Among the materials studied, Teflon favors the homeotropic alignment, while all other materials favor the planar alignment $\theta = \pi/2$. Similar plots for gold, silver, and copper are presented in figure~\ref{fig:oneseven}(b). All the metals favor the planar alignment. The absolute difference between the planar and homeotropic orientations is 4-6 zJ, i.e. almost an order of magnitude larger than for the dielectrics. 

For the planar surface alignment (as well as for any nonzero tilt angle), the azimuth orientation of the optical axes on the opposite sides of the slab can be different. It is of interest therefore to investigate the vdW free energy as a function of the azimuth misalignment $\psi \equiv \psi_3 - \psi_1$. Such dependencies are shown in figure~\ref{fig:twozero} for the planar alignment $\theta_1 = \theta_3 = \pi/2$. For all dielectrics and metals, the parallel orientation of the optical axes, $\psi_1 = \psi_3$, is energetically preferred to the cross-orientation $\psi_3 - \psi_1 = \pi/2$. However, this effect is relatively small, on the order of 0.1 zJ. Azimuth misalignment makes a small contribution of to the overall energy balance. It is probably smaller than the variation from the uncertainty in the material parameters.

\section{\label{sec:four}
Two parallel slabs
}

Two parallel slabs immersed in a liquid crystal are now considered. First, the spectrum of surfaces modes is derived from a product of four transfer matrices. Then the cases of semi-infinite slabs and finite-thickness slabs are analyzed in order.

\subsection{\label{sec:fourone}
General expression for the van der Waals energy
}

The system consists of five spatial regions separated by four interfaces, cf. figure~\ref{fig:one}. Collecting scattering at all four interfaces, the wave amplitudes in region {\bf 5} are expressed via the wave amplitudes in region {\bf 1} as follows
\begin{equation}
\left( \begin{array}{c} A_{\bf 5} \\  B_{\bf 5}  \end{array} \right) = 
\hat{M}_{\bf 54} \cdot \hat{M}_{\bf 43} \cdot \hat{M}_{\bf 32} \cdot \hat{M}_{\bf 21} 
\left( \begin{array}{c} A_{\bf 1} \\  B_{\bf 1} \end{array} \right) . 
\label{eq:nineteen}
\end{equation}
Surface states are defined by setting $A_{\bf 1} = B_{\bf 5} = 0$. The top equation (\ref{eq:nineteen}) defines the spatial structure of the surface mode. The bottom equation defines the spectrum. Developing the bottom equation via matrix elements one obtains
\begin{widetext}
\begin{equation}
\left( M_{\bf 54}^{21} M_{\bf 43}^{11} + M_{\bf 54}^{22} M_{\bf 43}^{21} \right) 
\left( M_{\bf 32}^{11} M_{\bf 21}^{12} + M_{\bf 32}^{12} M_{\bf 21}^{22} \right)  + 
\left( M_{\bf 54}^{21} M_{\bf 43}^{12} + M_{\bf 54}^{22} M_{\bf 43}^{22} \right) 
\left( M_{\bf 32}^{21} M_{\bf 21}^{12} + M_{\bf 32}^{22} M_{\bf 21}^{22} \right)  = 0 .  
\label{eq:twenty}
\end{equation}
Substituting here the explicit matrix elements from (\ref{eq:nine}) and (\ref{eq:ten}) and cancelling common positive-definite factors [this does not affect the final force after taking the logarithm in (\ref{eq:three})], the spectrum equation becomes
\begin{equation}
W = \left[ 1 - \Delta_{\bf 23} \Delta_{\bf 21} e^{- 2 q d_2} \right]  
    \left[ 1 - \Delta_{\bf 43} \Delta_{\bf 45} e^{- 2 q d_4} \right]        
  - \left[   \Delta_{\bf 23} - \Delta_{\bf 21} e^{- 2 q d_2} \right]  
    \left[   \Delta_{\bf 43} - \Delta_{\bf 45} e^{- 2 q d_4} \right]  
    \exp{\left( - \frac{2 p_{\bf 3} L}{\varepsilon^{zz}_{\bf 3}} \right) } = 0 \: .   
\label{eq:twentyone}
\end{equation}
Here $L = L_{32} - L_{43}$ is the gap between the slabs and the factors $\Delta_{\bf 43}$ and $\Delta_{\bf 45}$ are defined according to (\ref{eq:sixtyfour}) with {\bf 2} replaced by {\bf 4}. 

In accordance with the recipe (\ref{eq:three}), the free energy per unit interface area is 
\begin{equation}
F_1 = kT \sideset{}{'}\sum^{\infty}_{n = 0} \iint\limits_{-\infty}^{\makebox[0.5cm]{} \infty} 
\frac{{\rm d}q_x {\rm d}q_y}{(2\pi)^2} \ln \left[ W \right] . 
\label{eq:fifteen}
\end{equation}
If the gap between the slabs is large, $L \gg d_2, d_4$, then the second term in (\ref{eq:twentyone}) vanishes. The first term under the logarithm in (\ref{eq:fifteen}) splits into two parts, each corresponding to the vdW energy of an isolated slab surrounded by the medium. The total energy reduces to a sum of two terms derived in section~\ref{sec:seven}.    

If the slabs are thick, $d_2, d_4 \gg L$, the spectrum equation (\ref{eq:twentyone}) reduces to 
\begin{equation}
W^{\infty} = 1 - \Delta_{\bf 23} \Delta_{\bf 43} 
\exp{ \left( - \frac{2 p_{\bf 3} L}{\varepsilon^{zz}_{\bf 3}} \right) } = 0 \: . 
\label{eq:fifteenone}
\end{equation}
It will be analyzed later. Here the general expression (\ref{eq:fifteen}) is adapted for numerical evaluation. 

In the integral over $q$, polar coordinates $q_x = q \cos{\chi}$, $q_y = q \sin{\chi}$ are useful. Then the Hamaker ``constant'' $A = 12 \pi L^2 F_1$ is introduced to account for the usual $1/L^2$ scaling of the energy. It is also convenient to change the integration variable from $q$ to $t = qL$. The final expression for $A$ is
\begin{eqnarray}
A & = & 6kT \sideset{}{'}\sum^{\infty}_{n = 0} \int^{\infty}_{0} \!\! t \, {\rm d}t 
\int^{2\pi}_{0} \! \frac{{\rm d} \chi}{2\pi}  \times            
   \ln \left\{
   \left[ 1 - \Delta_{\bf 23} \Delta_{\bf 21} \exp{\left( - \frac{2d_2}{L} t \right)} \right]  \!
   \left[ 1 - \Delta_{\bf 43} \Delta_{\bf 45} \exp{\left( - \frac{2d_4}{L} t \right)} \right] \right.     
                                                                  \makebox[0.5cm]{}           
\nonumber \\
& & \makebox[2.0cm]{} \left. - \exp{ \left( - \frac{2 u_{\bf 3}}{\varepsilon^{zz}_{\bf 3}} t \right) }
         \left[   \Delta_{\bf 23} - \Delta_{\bf 21} \exp{\left( - \frac{2d_2}{L} t \right) } \right]  \!
         \left[   \Delta_{\bf 43} - \Delta_{\bf 45} \exp{\left( - \frac{2d_4}{L} t \right) } \right] 
\right\} \! .            
\label{eq:thirtyone}
\end{eqnarray}
\end{widetext}
The input parameters for numerical calculations are the geometrical factors $d_2/L$, $d_4/L$, the orientation of optical axes $\theta_{\bf r}$ and $\psi_{\bf r}$ and the temperature $T$. The last three parameters define the quantities $u_{\bf r}$ that enter via $\Delta_{\bf ij}$. 

In the limit of thin slabs the integrand is nonzero within the large interval $0 < t < L/d$ so the second term under the logarithm does not contribute much. Then $A$ converges to the quantity $R$ defined in section~\ref{sec:seven} multiplied by the factor $(L/d)^2$ that is responsible for the difference in definitions of $A$ and $R$.

\begin{figure*}[t]
\begin{center}
\includegraphics[width=0.98\textwidth]{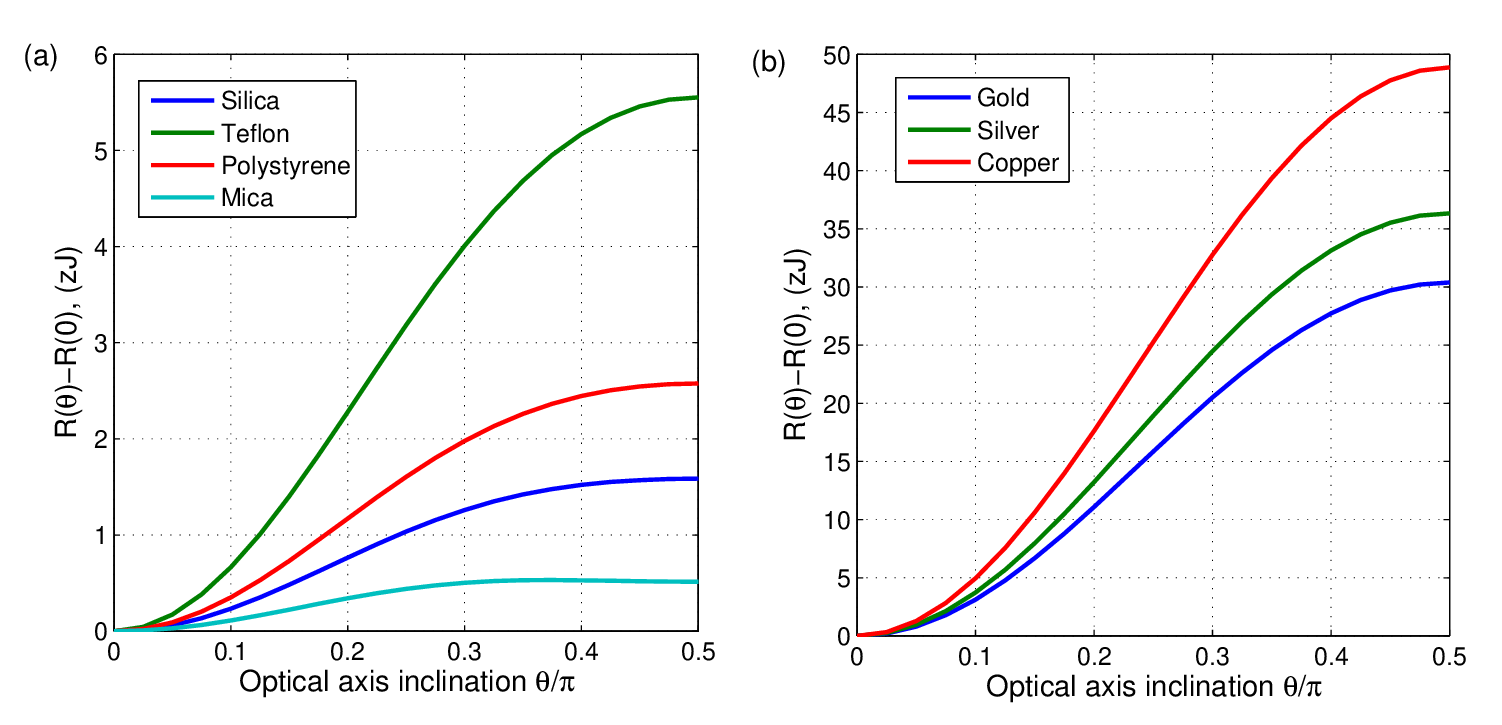}
\end{center}
\caption{(a) Hamaker constant (\ref{eq:eighteen}) for two semi-infinite bodies made of different dielectric materials separated by 5CB at $T = 298.2$ K. The reference values are $A^{\infty}(0) = -16.09$, $-16.57$, $-11.87$, and $-5.32$ zJ for silica, Teflon, polystyrene, and mica, respectively. The attraction is strongest at $\theta_3 = 0$. (b) Same for gold, silver, and copper separated by 5CB at $T = 298.2$ K. The reference values are $A^{\infty}(0) = -128.82$, $-144.94$, and $-283.55$ zJ.}
\label{fig:onefive}
\end{figure*}

\subsection{\label{sec:fourtwo}
Two semi-infinite slabs
}

Two semi-infinite bodies interacting via a gap $L$ is the basic vdW geometry. The spectrum of surface modes is given by (\ref{eq:fifteenone}). Going over to polar coordinates, expanding the logarithm and integrating over $q$ results in  
\begin{equation}
A^{\infty} = - \frac{3kT}{2} \sideset{}{'}\sum^{\infty}_{n = 0} 
\left[ \varepsilon^{zz}_{\bf 3} \right]^2
\sum^{\infty}_{m=1} \frac{1}{m^3} \int^{2\pi}_{0} \! \frac{{\rm d}\chi}{2\pi} 
\frac{ ( \Delta_{\bf 23} \Delta_{\bf 43})^m }{u^2_{\bf 3}} \: . 
\label{eq:eighteen}
\end{equation}

Figure~\ref{fig:onefive}(a) shows the Hamaker constant $A^{\infty}$ for several dielectric materials separated by 5CB, as a function of the tilt angle of liquid crystal's optical axis. For convenience of presentation, the Hamaker constants are referenced from their values at $\theta_3 = 0$. The reference values are listed in the caption. All the materials show preference of homeotropic alignment $\theta_3 = 0$. Among the materials studied, Teflon has shown the largest difference between the homeotropic and planar vdW energies (about 5.5 zJ).       

\begin{figure}[t]
\begin{center}
\includegraphics[width=0.48\textwidth]{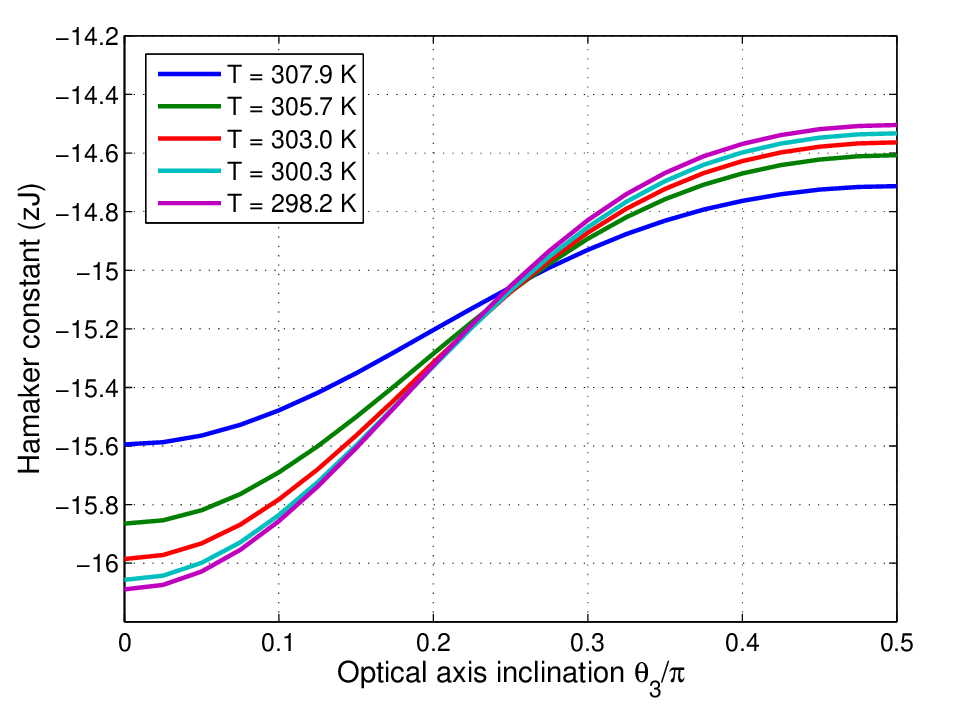}
\end{center}
\vspace{-0.5cm}
\caption{Hamaker constant (\ref{eq:eighteen}) for two semi-infinite silica bodies separated by 5CB for five temperatures.}
\label{fig:three}
\end{figure}

Metals possess qualitatively similar angle variations of vdW energy, as shown in figure~\ref{fig:onefive}(b). The attraction is stronger for the homeotropic alignment. However, the absolute scale of the variation is about one order of magnitude larger than for dielectric materials. A stronger vdW attraction for homeotropic alignment may be a general feature of 5CB and perhaps of any positive liquid crystal.  
     
One expects the inclination dependence to go away as the temperature increases and the medium becomes optically isotropic. The temperature dependence of $A^{\infty}(\theta_3)$ for the silica-5CB system is shown in figure~\ref{fig:three}.

\subsection{\label{sec:fourthree}
Two finite-thickness slabs
}

To analyze a system of two slabs oriented parallel to each other, the full four-transfer matrix solution (\ref{eq:thirtyone}) is needed. The present study is focused on finding an optimal orientation of the optical axes that minimizes the vdW free energy. At large separations $L \gg d_2, d_4$, one expects one-slab effects to be dominant. For all materials except Teflon the vdW energy is minimal when the liquid crystal is planar aligned on both sides of the slab. In the opposite limit of very small separations, $L \ll d_2, d_4$, one expects the interaction effects to dominate. As discussed above the interaction energy favors the homeotropic alignment for all the materials studied. It is of interest therefore to study the evolution of the optimal orientation of the middle liquid crystal and to follow the interplay between the single-slab and interaction effects. 

\begin{figure}[t]
\begin{center}
\includegraphics[width=0.48\textwidth]{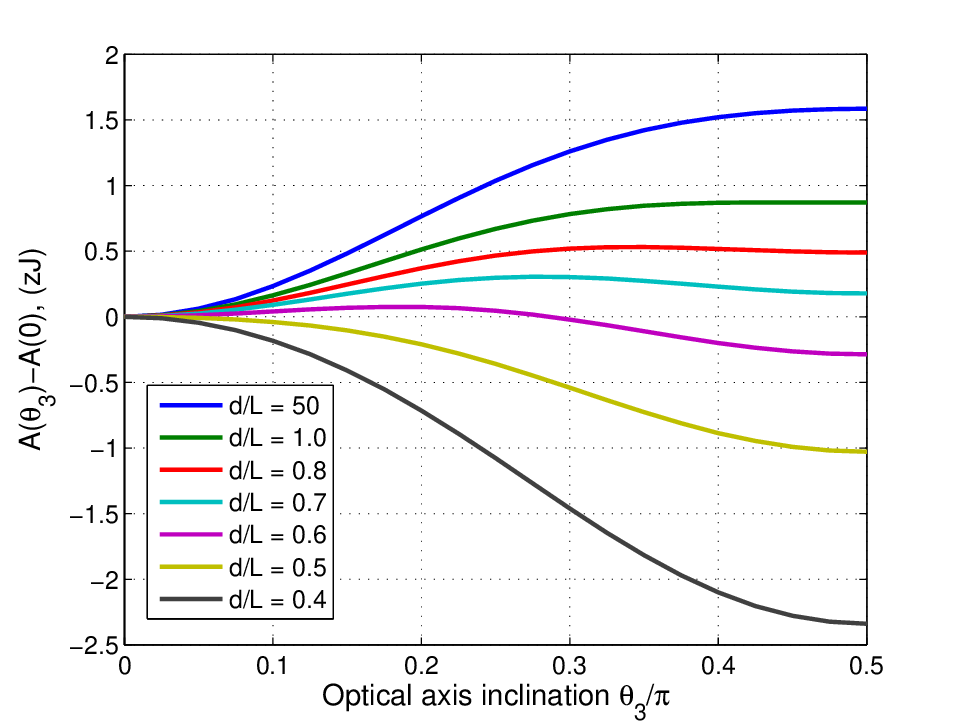}
\end{center}
\caption{Hamaker ``constant'' $A$, (\ref{eq:thirtyone}), of two silica slabs in 5CB for several slab thicknesses. The alignment of liquid crystal on the outside surfaces is planar, $\theta_1 = \theta_5 = \pi/2$, while the alignment in the gap is varied. The reference values are $-16.09$, $-39.21$, $-54.25$, $-67.33$, $-87.77$, $-122.12$, $-186.06$ zJ, (for $d/L$ going from large to small). The absolute temperature is $T = 298.2$ K.}
\label{fig:twoone}
\end{figure}

Consider the 5CB-silica-5CB-silica-5CB system as an example. A single silica slab in 5CB favors the planar alignment on both surfaces. Accordingly, one sets $\theta_1 = \theta_5 = \pi/2$, and $\phi_1 = \psi_3 = \psi_3 = 0$. The tilt angle of the middle LC section $\theta_3$ remains variable to include the possibility of homeotropic and other alignments. The slabs are assumed to be of the same thickness, $d_2 = d_4 \equiv d$. The overall vdW energy is studied as a function of $\theta_3$ for different ratios $d/L$.         

Results of numerical calculations are shown in figure~\ref{fig:twoone}. For large $d/L > 5$, the vdW energy is dominated by the interaction across the gap, and the $\theta_3$ dependence is virtually the same as in the semi-infinite case. (Compare the $d/L = 50$ plot in figure~\ref{fig:twoone} with the $T = 298.2$ K plot in figure~\ref{fig:three}.) As the slabs get thinner, the single-slab effects grow stronger and eventually dominate. Accordingly, the planar orientation $\theta_3 = \pi/2$ becomes the absolute energy minimum at $d/L \leq 0.5$. At intermediate thicknesses, $0.5 < d/L < 1.0$, the energy has two local minima, at $\theta_3 = 0$ and $\theta_3 = \pi/2$, as can be seen in the figure.

\section{\label{sec:nine}
Summary 
}

Collective behaviour of colloidal particles in aniso\-tro\-pic media is a fascinating and complex subject. This behaviour is determined by the balance of surface alignment energy, bulk elastic energy, electrostatic forces, van der Waals forces, and thermal fluctuations. Given the technological importance of both liquid crystals and non-aqueous colloids it would be desirable to have a comprehensive theory of colloidal stability in liquid crystals of the same clarity as the classical theory of colloidal stability in isotropic fluids~\cite{Verwey1999}. In recent years, significant progress has been made in understanding the elastic contribution~\cite{Lubensky1998,Stark2001,Eskandari2012}. 

The main purpose of the present work has been to demonstrate that another component in this mix, the van der Waals force, becomes anisotropic. Due to the complexity of the general problem, only the plane geometry in the nonretarded limit has been analyzed. The vdW free energy can be obtained in this case from the spectrum of electromagnetic surface modes relatively easily. Unlike previous works, the focus here has been the dependence of vdW energy on the optical axis inclination angle.   

A significant barrier for any realistic calculation of vdW forces is the lack of reliable parameterization of the dielectric function over the entire imaginary frequency axis. In liquid crystals, this is further complicated by birefringence and strong temperature dependence. In this paper, a three-oscillator temperature-dependent model of 5CB has been constructed based on the real-frequency data of Wu and co-workers~\cite{Wu1990,Wu1991,Wu1993,Li2004}. More work will be needed to further validate and refine the model presented in section~\ref{sec:five}.      

In an anisotropic fluid the vdW energy of a parallel-plate slab becomes a function of the tilt angle between the optical axis and the surface normal. Energy profiles have been calculated in section~\ref{sec:seven}. All studied materials except Teflon favor the planar alignment of the optical axis. If the real anchoring orientation is different from the optimal one, the dependence of the vdW energy on the tilt angle will generate a vdW torque that needs to be taken into consideration in determining the equilibrium orientation of the slab. The torque disappears as temperature is raised above the nematic-isotropic transition.   

In the case of planar alignment and other non-ho\-me\-o\-tro\-pic surface alignments, the optical axes may have different azimuth orientation on either side of the slab. The vdW energy is in general a function of the azimuth angle mismatch. This effect has been investigated in section~\ref{sec:seven} and found to be numerically small. All the materials studied favor parallel orientation of the optical axes, i.e. equal azimuth angles on both surfaces.     

When two slabs are brought close together, they attract via a vdW force that is a function of the optical axis direction of the intermediate medium. A general solution to this problem has been developed in section~\ref{sec:fourone}. It has been found that the vdW force is strongest for the homeotropic orientation of 5CB for all the materials analyzed. 5CB is a positive liquid crystal. The main result suggests the vdW attraction is strongest when the surface normal is parallel to the line of largest polarizability of the medium molecules. 

The effects of finite slab thickness have been investigated. Using the general solution (\ref{eq:thirtyone}) smooth evolution of the vdW energy from the gap dominated limit to the slab-thickness dominated limit has been observed. At least for some materials (such as silica in 5CB) this implies that the planar alignment in the gap is preferred for thin slabs and large gaps, while the homeotropic alignment is preferred for thick slabs and small gaps.    

One should add in closing that these conclusions cannot be directly applied to real colloidal suspensions without first including the elastic and electrostatic contributions.

\begin{acknowledgments}

This work grew out of a project at Hewlett-Packard on the dynamics of charged colloids in nonaqueous solvents. The author wishes to thank Susanne Klein and Vladek Kasperchik for illuminating discussions on the subject of this paper. 

\end{acknowledgments}

\end{document}